# Observation of Magnetic Radial Vortex Nucleation in a Multilayer Stack with Tunable Anisotropy

Vedat Karakas[1*], Aisha Gokce[1], Ali Taha Habiboglu[1], Sevdenur Arpaci[1], Kaan Ozbozduman[1], Ibrahim Cinar[1,2], Cenk Yanik[3], Riccardo Tomasello[4], Silvia Tacchi[5], Giulio Siracusano[6], Mario Carpentieri[7], Giovanni Finocchio[6], Thomas Hauet[8], Ozhan Ozatay[1]

[1]Physics Department, Bogazici University, Bebek 34342, Istanbul, TURKEY

[2]Karamanoglu Mehmetbey University, Department of Physics, 70100 Karaman, TURKEY

[3]Sabanci University Nanotechnology Research and Application Center Tuzla, 34956 Istanbul, TURKEY

[4]Department of Engineering, Polo Scientifico Didattico di Terni, University of Perugia, Terni, ITALY

[5]Istituto Officina dei Materiali del CNR (CNR-IOM), Sede Secondaria di Perugia, c/o Dipartimento di Fisica e Geologia, Università di Perugia, Perugia, ITALY

[6]Department of Mathematical and Computer Sciences, Physical Sciences and Earth Sciences, University of Messina, Messina, ITALY

[7]Department of Electrical and Information Engineering, Politecnico di Bari, I-70125 Bari, ITALY

[8]Institut Jean Lamour, UMR CNRS-Université de Lorraine, 54506 Vandoeuvrelès Nancy, FRANCE

Recently discovered exotic magnetic configurations, namely magnetic solitons appearing in the presence of bulk or interfacial Dzyaloshinskii-Moriya Interaction (i-DMI), have excited scientists to explore their potential applications in emerging spintronic technologies such as race-track magnetic memory, spin logic, radio frequency nano-oscillators and sensors. Such studies are motivated by their foreseeable advantages over conventional micro-magnetic structures due to their small size, topological stability and easy spin-torque driven manipulation with much lower threshold current densities giving way to improved storage capacity, and faster operation with efficient use of energy. In this work, we show that in the presence of i-DMI in Pt/CoFeB/Ti multilayers by tuning the magnetic anisotropy (both in-plane and perpendicular-to-plane) via interface engineering and postproduction treatments, we can stabilize a variety of magnetic configurations such as Néel skyrmions, horseshoes and most importantly for the first time, the recently predicted isolated radial vortices at room temperature and under zero bias field. Especially, the radial vortex state with its absolute convergence to or divergence from a single point can potentially offer exciting new applications such as particle trapping/detrapping in addition to magnetoresistive memories with efficient switching, where the radial vortex state can act as a source of spin-polarized current with radial polarization.

**Introduction**

Magnetic skyrmions are spin configurations with a topology that has perpendicular-to-plane magnetization components at the core and the edges with opposite directions[1,2]. They can be Bloch or Néel type depending on the chirality of the transition region between the core and the edges, being circular or radial, respectively[3]. Unique properties of skyrmions such as their intrinsically small size, topological stability and efficient manipulation with much lower threshold current densities compared to conventional micromagnetic structures have recently attracted the attention of researchers to look for ways of utilizing them in technological applications. Envisioned skyrmionic devices[1,2] are expected to possess the benefits of combining storage, logic operations and microwave functionalities at the same level with efficient use of energy[4,5].

Skyrmions appear due to Dzyaloshinskii-Moriya Interaction (DMI) in the bulk of chiral magnets (Bulk DMI), at the interface of heavy metal/ferromagnet thin film stacks (interfacial DMI)[6–8] or in perpendicular magnetic anisotropy materials as a result of long range dipolar interactions[9,10] in the presence of DMI. Bulk DMI arises as a result of lack of inversion symmetry in chiral magnets, whereas the interfacial DMI (i-DMI) stems from the interaction between ferromagnetic atoms and strong spin-orbit coupling (SOC) atoms of an adjacent heavy metal[11–13]. I-DMI strength is parameterized by a constant D and can be incorporated into the Landau-Lifshitz-Gilbert (LLG) equation competing with other energy terms such as exchange, anisotropy and magneto-static energies. The resulting micromagnetic configuration in systems with i-DMI can lead to a range of interesting canted spin orientations such as Néel skyrmions[14], horseshoes[15], spider-web domains[3] and radial vortices[3].

In this paper, we demonstrate an evolution of these spin structures in 60 and 15 repeat $Pt(5)/Co_{20}Fe_{60}B_{20}(1)/Ti(1)$ multilayers (all thicknesses in nm), partially and fully patterned into nanopillar disks, respectively. The disk diameter, i-DMI strength D and magnetic anisotropy play the prominent role in determining the resulting magnetic configuration. An advantage of using repeated magnetic multilayer stacks is the enhancement of the magnetic signal and the interface quality with increasing number of repeats. Most importantly, we were able to tune the magnetic anisotropy by adjusting the number of repeats in the thin film stack as well as performing a set of post-production treatments such as ion irradiation during milling and annealing, in order to be able to stabilize both radial vortices and Néel skyrmions at room temperature. In fact, we were able to detect these solitons via Magnetic Force Microscopy (MFM) with an in situ parallel to the thin film surface plane magnetic field application capability. This technique serves as a quick, straightforward and cost effective method unlike conventional skyrmion imaging techniques such as Lorentz Transmission Electron Microscopy[11], Scanning Tunneling Microscopy[6] or Small Angle Neutron Scattering[13] at the cost of a reduced spatial resolution limited by the magnetic tip diameter (on the order of 50 nm in our case).

Heavy-Metal/Ferromagnet/Spacer-Metal multilayer stack (for our case $[Pt(5)/Co_{20}Fe_{60}B_{20}(1)/Ti(1)]_n$) with variable number of repeats n, patterned into a nanostructure provides a unique workbench for the study of various micromagnetic structures nucleated within both in-plane and perpendicular-to-plane anisotropies in the presence of i-DMI. The ability to investigate exotic magnetic states with i-DMI and in-plane anisotropy allowed us to discover for the first time the nucleation of recently predicted radial vortices that are stable under zero bias field and at room temperature. These radial vortices can potentially offer novel applications such as particle trapping/detrapping by applying a magnetic field with radial symmetry in addition to non-uniform magnetization magnetoresistive memory where radial vortex chirality is fixed for one layer and free to switch for the other, providing full read out signal. The radial vortex can also act as a source of spin-polarized current with radial polarization for spin-torque memory and microwave nano-oscillator devices.

**Experimental Measurements**

The experimental studies were performed on $[Pt(5)/Co_{20}Fe_{60}B_{20}(1)/Ti(1)]_n$ multilayers (all thicknesses in nm) with number of repeats n being 1, 10, 15, and 60. Material stack, including a 5 nm Ti adhesion layer and a 5 nm Pt capping layer, was grown on $Si/SiO_2$(500 nm) substrate via DC magnetron sputtering in 2 mTorr Ar pressure starting from a base pressure of ~ $10^{-9}$ Torr. (For more details see methods.)

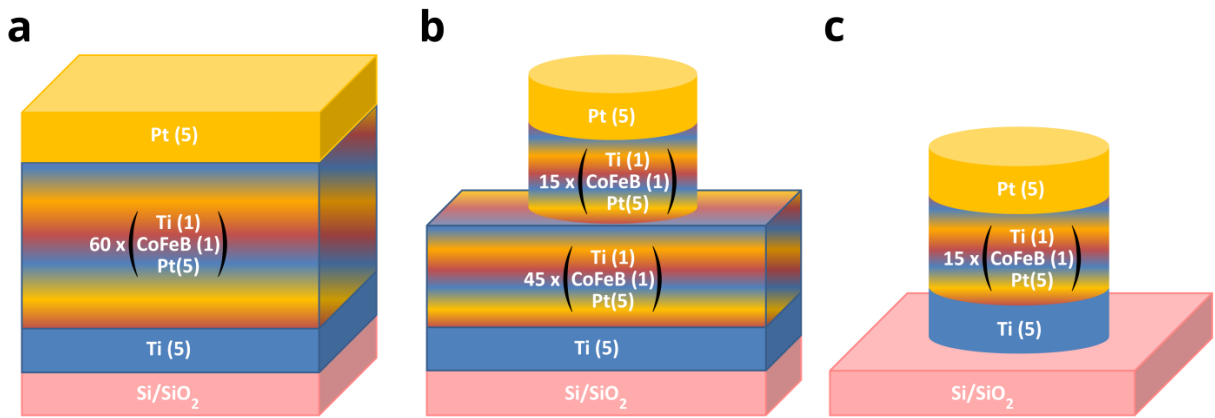

**Figure 1  Schematic of three sample types.** a) 60 repeat continuous film, sample 60C, b) 15 repeat partially patterned, sample 15PP-A c) 15 repeat fully patterned, sample 15FP-AI. Sample label conventions are defined in Table 1.

| Name | # of Repeats | Measurement Technique | Continuous Film | Partially Patterned | Completely Patterned | Post-production Treatment ||| 
|------|-------------|----------------------|-----------------|--------------------|--------------------|-----|-----|-----|
|      |             |                      |                 |                    |                    | None (As deposited) | Annealed | Ion-irradiated |
| 1C | 1 | BLS | X | | | X | | |
| 10C | 10 | BLS | X | | | X | | |
| 15C | 15 | VSM | X | | | X | | |
| 15C-AI | 15 | VSM, XRD and MFM | X | | | | X | X |
| 60C | 60 | VSM and XRD | X | | | X | | |
| 15PP-A | 45 extend. 15 pattern. | MFM (nanodisk) | | X | | | X | |
| 15FP-AI | 15 pattern. | MFM (nanodisk) | | | X | | X | X |

**Table 1: Production parameters and measurement techniques for individual samples.** Sample naming convention: [# of repeats][Thin film property (continuous / fully or partially patterned)]-[Post production treatments (annealing and/or ion irradiated)]. Abbreviations: C for continuous, A for annealed, I for ion-irradiated during milling, PP for partially patterned, FP for fully patterned

Table 1 uses the sample label convention where the number of repeats in multilayer stack is 1, 10, 15 or 60, C stands for continuous films as shown in Fig. 1a. Post production treatment A stands for annealed at 170ºC for 45 minutes and I for ion-irradiation during milling with 300 V beam voltage, 60 V Accelerator voltage, 12 mA Beam current for 2 hours. The sample 15PP-A consists of 45 repeat continuous film underneath the 15 repeat nano disks as shown in Fig. 1b and the sample 15FP-AI, obtained by ion mill etching of 45 repeats of sample 15PP-A, is completely patterned into 15 repeat nano disks as shown in Fig. 1c. The measurement techniques are Brillouin Light Scattering (BLS), Vibrating Sample Magnetometer (VSM), X-ray diffraction (XRD) and Magnetic Force Microscopy (MFM).

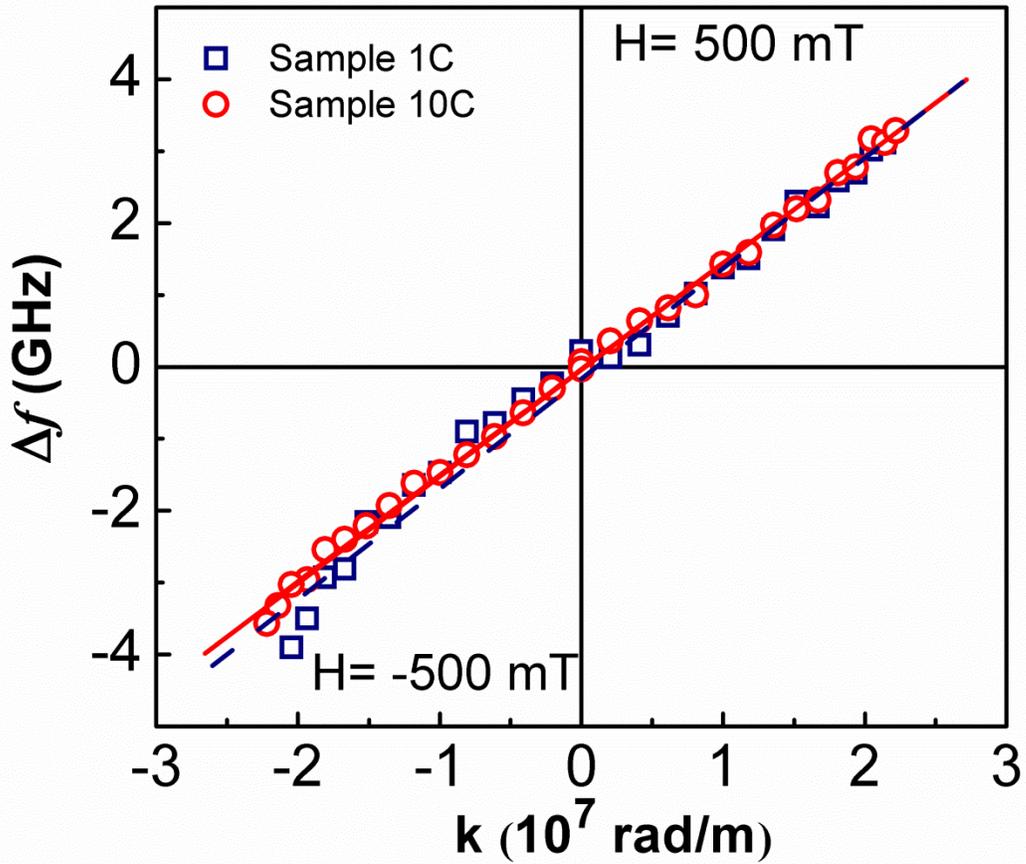

**Figure 2 BLS measurements of the full film samples.** Red circle for 10 repeat (10C) sample and blue square for single repeat sample (1C). Frequency difference, $\Delta f$, between Stokes and anti-Stokes peaks is plotted against wave vector k for 500 mT bias field applied parallel to the film plane; k values are inverted for negative bias field -500 mT. Line fits to the data of samples 10C and 1C are shown by a red line and a blue dashed line, respectively.

Brillouin Light Scattering (BLS) has been exploited to evaluate the i-DMI strength D originating at the interface of Pt/CoFeB for a single repeat, 1C, and 10 repeat stack, 10C. In ultrathin films, the presence of i-DMI induces a noticeable frequency asymmetry between oppositely propagating Damon-Eshbach (DE) modes, which increases as a function of the wave vector k following the relation[16]:

$$\Delta f = f_{DMI}(k) - f_{DMI}(-k) = \frac{2\gamma D}{\pi M_s} k \qquad (1)$$

where $\gamma$ is the gyromagnetic ratio[17], and $M_s$ is the saturation magnetization of the magnetic layer. Using BLS the frequency asymmetry caused by i-DMI can be quantified by measuring the frequency difference, $\Delta f$, between Stokes and anti-Stokes peaks corresponding to spin waves propagating in opposite directions. Fig. 2 shows $\Delta f$ measured (points) as a function of wave vector k for both samples, applying a bias field of +500 mT and -500 mT. The strength of i-DMI has been determined by a linear fit of the experimental data, using equation (1). The D values were found to be 1.49±0.05 mJ/m$^2$ and 1.45±0.02 mJ/m$^2$, for single repeat (1C) and 10 repeat stack (10C) samples, respectively. It is known that the D value, for a fixed CoFeB thickness of 2 nm, saturates[16] at 0.45 mJ/m$^2$ for Pt thicknesses greater than 2 nm whereas for a CoFeB thickness of 0.8 nm, D increases[17] to 1.0±0.1 mJ/m$^2$. In our samples characterized by a CoFeB thickness of 1 nm and a Pt thickness of 5 nm (greater than the saturation thickness of 2 nm), we found D values of 1.4 - 1.5 mJ/m$^2$, which are larger than the previously reported[17] value of 1.0±0.1 mJ/m$^2$. These higher D values can be ascribed to the slightly different Co/Fe ratio of 20/60 as compared to 40/40 in refs. 18, 19 and different deposition conditions. Remarkably, we found that for [Pt(5)/Co$_{20}$Fe$_{60}$B$_{20}$(1)/Ti(1)]$_n$ stacks where n>1, the presence of a non-heavy metal Ti interlayer 1 nm thick, successfully avoids degradation[7,18] of i-DMI strength due to symmetric contribution of bottom and top Pt electrodes. The fact that D saturates as a function of number of repeats (approximately the same D values for sample 1C and sample 10C) implies that Ti interlayer tends to delete the spin-orbit coupling from the top Pt.

Fig. 3a shows room temperature, in-plane field Vibrating Sample Magnetometer (VSM) measurements of 15 and 60 repeats of the full film stack [Pt(5)/Co$_{20}$Fe$_{60}$B$_{20}$(1)/Ti(1)]$_n$ (as deposited samples 15C, 60C and with the post-deposition treatment of annealing at 170 °C for 45 minutes and ion-irradiation during milling for 2 hours for sample 15C-AI). The post deposition treatments were done to clarify the role of nano patterning of resist bake and ion mill etch procedures on the resulting device performance.

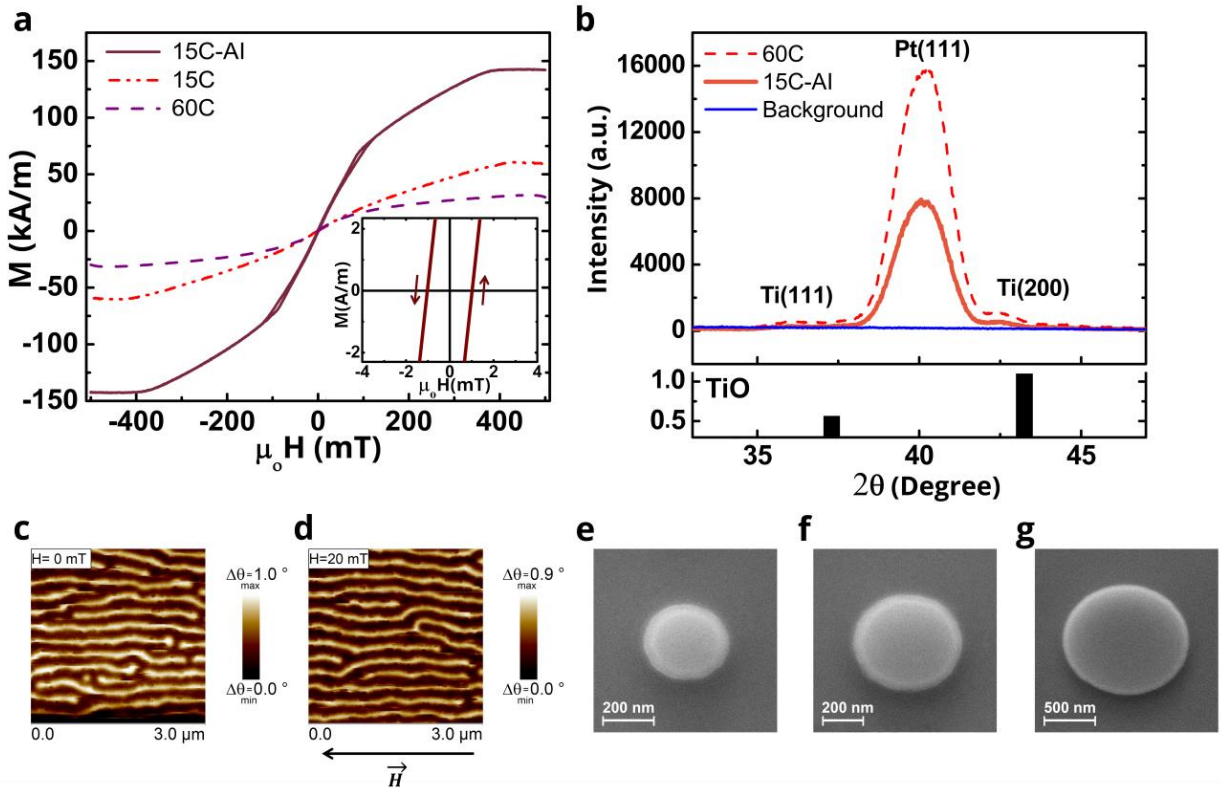

**Figure 3** a) Room temperature VSM measurements with in-plane external magnetic field for samples 15C (red dash-dotted line), 15C-AI (brown-red solid line) and 60C (purple dashed line). The inset shows the zoomed in version for sample 15C-AI, the only one showing hysteresis. The arrows indicate the field scan direction. b) XRD data for samples 15C-AI (orange solid line) and 60C (red dashed line) are shown along with Si/SiO$_2$ substrate data as background (blue solid line). The crystal orientations corresponding to the diffraction peaks are identified as being due to Pt[19] and Ti that gets progressively oxidized to TiO thin films suggesting a strong <111> texture. For reference JCPDS card data of TiO is included[20]. c-d) MFM images of sample 15C-AI after saturation with 1 T in-plane field, (c) at remanence, (d) under 20 mT in-plane field applied along the direction indicated by the arrow. e-g) Scanning Electron Microscopy (SEM) images of nano disk with various diameters, e) 300 nm f) 500 nm and g) 1200 nm.

The first trend evident in Fig. 3a is the reduction of magnetization, at an applied field of 500 mT, from 60 kA/m for sample 15C to 31 kA/m for sample 60C as a result of the increase in number of repeats. The linear M-H curve suggests an in-plane hard axis, *i.e.* an out-of-plane anisotropy which gets stronger as the number of repeats increase. We estimated a perpendicular anisotropy constant $K=2.1 \times 10^4$ J/m$^3$ for sample 60C and $K=1.3 \times 10^4$ J/m$^3$ for sample 15C, whereas we found that postproduction treatment of annealing and ion irradiation during milling switched the easy axis direction towards in-plane for sample 15C-AI with in-plane anisotropy constant $K=1 \times 10^4$ J/m$^3$. Furthermore, the magnetization at 500 mT for sample 15C-AI increased to 141 kA/m as compared to 15C and 60C. The inset of Fig. 3a shows a zoomed in version of the data for sample 15C-AI. Unlike the other two samples 15C and 60C, sample 15C-AI displays a clear hysteresis with 1 mT coercive field.

In order to gain more insights into the behavior of sample 15C-AI, we investigated the magnetic configuration under Magnetic Force Microscopy (MFM) with a custom-made magnetic stage that has in-plane field application capability, after saturation with 1 T in-plane field. Fig. 3c and Fig. 3d show the MFM images of the remnant state and in the presence of 20 mT in-plane field, respectively. The effect of increasing the in-plane magnetic field from 0 to 20 mT was the growth of average size of the dark domains from 122 nm to 135 nm (See Fig. S2 for average domain size estimation details). Furthermore, domain walls were observed to move incrementally as a function of increasing field consistent with low field behavior seen in the VSM measurement Fig. 3a inset. We attribute this gradual change in the domain sizes and accompanying incremental domain wall motion to the presence of defects acting as pinning centers. Most importantly, the formation of the stripe domains confirm that the post-production treatment causes a reduction of the out-of-plane magnetic anisotropy and consequently the rotation of the easy axis from the perpendicular direction towards the surface plane.

To understand the role of number of repeats and post-production treatments on the anisotropy direction, X-ray diffraction (XRD) measurements with Cu K$_\alpha$ radiation were performed on 60C and 15C-AI samples. The results are shown in Fig. 3b which also displays the Si/SiO$_2$ substrate background data. Thin film crystal orientations were identified by the corresponding 2θ values of each individual peak. Highest intensity peaks of both 15C-AI and 60C occurred at 40.15° which indicate preferred Pt <111> crystal orientation[19] while the peaks appearing at 42.3° and 36.1° were identified as Ti <200> and Ti <111> peaks. Pt <111> peak intensity was doubled for 60C as compared to 15C-AI. We observed the presence of TiO peaks in accordance with Joint Committee on Powder Diffraction Standards (JCPDS) card data[20]. This behavior can be understood considering the fact that the substrate temperature gradually increases from 20 °C to 60 °C in the course of 4-hour-long deposition of the multilayer stack promoting further oxidation of Ti interlayers (See Table S1 and Fig. S5), which indicates that the ratio of TiO to Ti increases with number of repeats. From literature[21], lattice constants of Ti, TiO and Pt can be found as 4.68 Å, 4.18 Å and 3.92 Å, respectively, corresponding to a decreasing lattice mismatch from 16% to 6% with increasing oxygen content. The decreasing lattice strain favors strong Pt <111> orientation and the perpendicular magnetic anisotropy (PMA) induced by this crystal structure in sample 60C whereas the increasing magneto-elastic energy due to larger lattice strain alleviates PMA in sample 15C-AI. This also explains why the PMA of sample 15C is much weaker than sample 60C. Enhancement of the quality of CoFeB/Pt interface and Ti/Pt interface results in modification of anisotropy as evidenced by the evolution of spectrum of possible micromagnetic configurations[3], see Fig. 4 and Fig. 5.

Following the BLS measurements of i-DMI strength, magnetic characterization with VSM and structural characterization with XRD, top down method involving electron beam lithography and ion mill etching were utilized to pattern the films into nano disks with diameters ranging from 100 to 1200 nm. Figure 3e-g show SEM images of 300, 500 and 1200 nm diameter nanopillar disks that are of interest in our MFM studies.

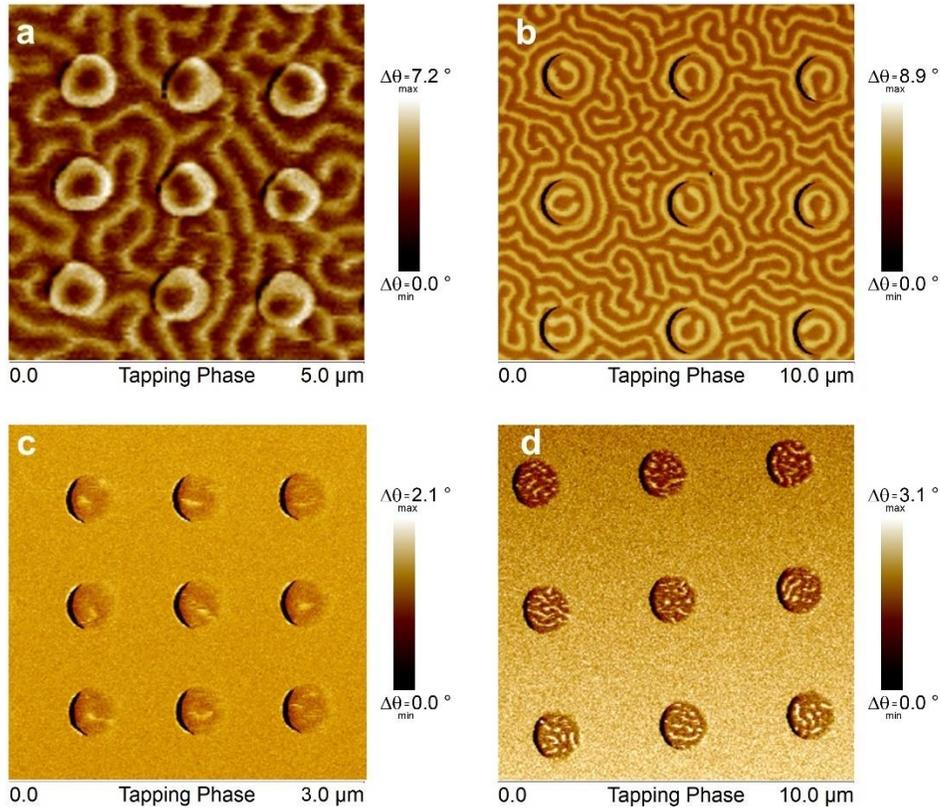

**Figure 4 MFM images of disk arrays** for 15PP-A (a-b) and 15FP-AI (c-d) samples taken under no external field after in-plane AC demagnetization with 1 T maximum field and application of 3 T perpendicular-to-plane field. MFM images belong to arrays of disks with diameters a) 500 nm, c) 300 nm, b) and d) 1200 nm.

Figure 4a and Fig. 4b show MFM images of the sample type 15PP-A for 500 nm and 1200 nm diameter disks, respectively. The samples were ac-demagnetized with a maximum in-plane field of 1 T followed by the application of a 3 T out-of-plane field, prior to imaging. The continuous film underneath the pillars was observed to have a maze-like domain pattern as expected due to PMA in these samples. The i-DMI at the CoFeB/Pt interface together with PMA leads to stabilization of Néel skyrmions in 500 nm disks (which is reconfirmed by their response to out-of-plane external field as predicted, see Supplementary Notes, Fig. S4) as shown in Fig. 4a and horse-shoe domain patterns in 1200 nm disks of Fig. 4b.

MFM images in Fig. 4c and Fig. 4d were obtained on 15FP-AI type samples having in plane anisotropy (IPA) after AC demagnetization process with a maximum in-plane field of 1 T and a subsequent 3 T out-of-plane magnetic field application. 1200 nm disks of Fig. 4d show a multi-domain pattern including stripe-like domains reminiscent of a spider web configuration[3]. On the other hand, 300 nm disks display stable exotic magnetic configurations that appear as dots or extended dots as seen in Fig. 4c. It is rather difficult to identify the nature of the magnetic ordering in this case since there are multiple possible magnetic structures such as Néel skyrmions or magnetic vortices that show similar contrast in an MFM image.

One way to diagnose the type of ordering is to analyze the characteristic behavior under the effect of an external magnetic field[3]. The possibility of having Néel skyrmions is eliminated by the fact that these samples (15FP-AI) have in-plane anisotropy and under 15 mT perpendicular-to-plane field they expand as predicted[22] (see supplementary notes Fig. S4). The other two candidates, namely circular vortices and radial vortices, both of which are stable in samples with in-plane anisotropy can luckily be easily distinguished by their response to external magnetic field. If they are circular vortices, they are expected to move in a direction perpendicular to the field whereas the recently predicted radial vortices are expected to move parallel or antiparallel to the applied field depending on the radial chirality coupled with core polarity (radially inward - core into the plane for the former or radially outward - core out of plane for the latter)[3].

Results of our external field dependent response analysis are summarized in Fig. 5. The MFM data belong to a single array of 300 nm diameter disks that went through two different magnetic field conditioning procedures. Figure 5a shows the zero external field configuration of the array after an in-plane AC demagnetization process of 1 T maximum field and a subsequent 3 T out-of-plane field application. Majority of the disks display the dot and extended dot structures. Due to the finite diameter of 50 nm for the low moment magnetic tip used in this imaging, the minimum feature size that can be resolved is limited to 50 nm. Therefore, the dot sizes we observed in the range of 50 to 100 nm merely imply an upper bound of 100 nm dot diameter.

Figure 5b shows the same disks imaged under an in-plane magnetic field of 10 mT along the direction indicated in the figure. The response to the external field is highlighted such that an upward or downward motion (parallel or anti-parallel to the field direction) is indicated by triangles, appearance of the dots in disks after field application is denoted by a − sign and disappearance of the dots is represented by an x.

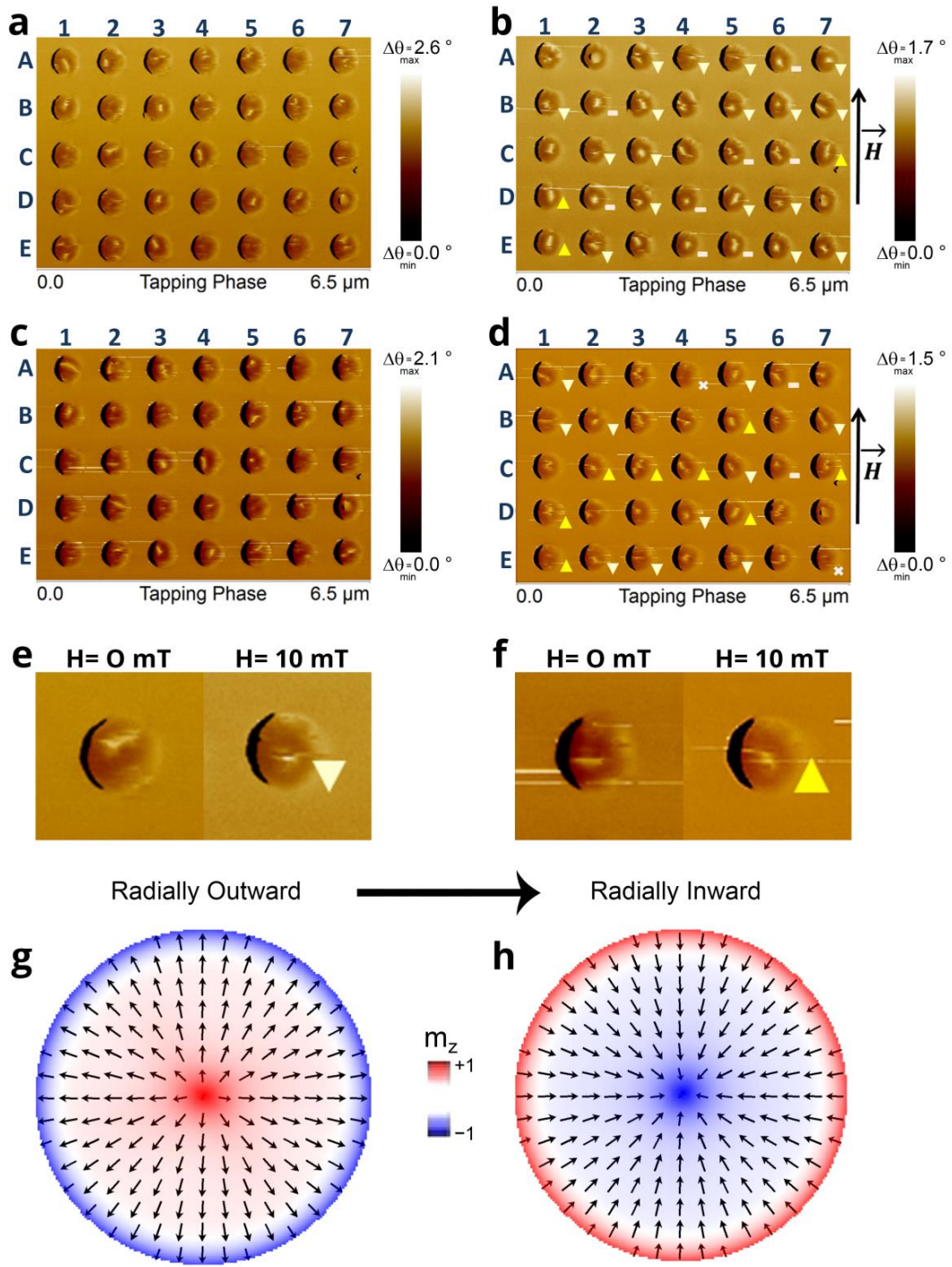

**Figure 5** MFM images of a single array of 300 nm diameter disks sample type 15FP-AI under different conditions. a-b) 3 T out-of-plane field and c-d) -4 T out-of-plane field, both following an AC demagnetization process with a maximum 1 T in-plane field. a and c MFM images taken with no external field, b and d under an external in-plane field of 10 mT applied in the direction indicated by the arrows. The response to the external field is highlighted such that an upward or downward motion is indicated by triangles, appearance of the dots in disks after field application is denoted by a – sign and disappearance of the dots is represented by an x. Zoomed view of the nanopillar disk located at C2 on the array is shown, after e) 3 T and f) -4 T out-of-plane field treatments with (right) and without (left) 10 mT in plane external field Micromagnetic simulation result obtained by using the experimental parameters indicate radially g) outward h) inward radial vortices.

23% of the disks showed no clearly resolvable features in Fig. 5a but revealed dots when the field was applied in Fig. 5b; e.g. the disk A6. One plausible explanation is that the diameters of the dots are much smaller than the resolution of the MFM tip and seem to be visible under the application of an external magnetic field. The striking behavior in Fig. 5b was that 49% of the dots moved downward, opposite to the external field whereas 9% moved upward along the field direction. This motion was either in the form of the dot relocating such as in disk E6 or extending in the vertical direction as in B3 and C7, the latter possibly due to pinning[22] effects which also explains how for example in disk A6 a dot becomes resolvable after the field application. For the remaining 19% of the dots, the behavior was not clear. See Supplementary Information and Fig. S1 for the analysis of a larger array including 90 nanopillar disks. It is evident from this experiment that majority of the dot-like structures move either parallel or anti-parallel to the direction of the field. These results support the idea that the observed magnetic features are radial vortices.

Now, to look for further evidence, we studied the possibility of nucleating radial vortices with a switched chirality. Figures 5c and 5d display the results of the same experiment repeated for the disks that were magnetized with -4 T out-of-plane magnetic field after an in-plane AC demagnetization process with a maximum 1 T in-plane field. Some of the dots that previously seemed to move antiparallel to the field direction in the presence of a 10 mT field (the expected response of a radial vortex with radially inward chirality) changed their behavior after -4 T out-of-plane field application and started to move parallel to the field direction (as would be expected for a radial vortex with radially outward chirality) as can be seen for example for nanopillar disk located at C2 as shown in Fig.5 e-f. Therefore, not only the radial vortex expected behavior is reproduced but also the possibility of nucleating radial vortices with a switched chirality is clearly demonstrated. Furthermore, given the experimental parameters, micromagnetic simulations confirm the stabilization of radial vortex structure with either outward (Fig. 5g) or inward (Fig. 5h) radial chirality on nanopillar disks.

**Conclusions and Discussion**

In summary, a variety of magnetic structures were observed in nano-disks made of alternating layers of Pt(5)/Co$_{20}$Fe$_{60}$B$_{20}$(1)/Ti(1) with i-DMI at the interface of the heavy metal Pt and the magnetic Co$_{20}$Fe$_{60}$B$_{20}$ layer. Previous studies report an i-DMI parameter of ~1 mJ/m$^2$ for Pt/Co$_{40}$Fe$_{40}$B$_{20}$ interface[17,23] and ~2 mJ/m$^2$ for Ir/Co/Pt heterostructure that has two heavy metal layers[7] having opposite i-DMI signs to prevent the cancellation in repeated structure. From BLS analysis we found that Pt(5)/Co$_{20}$Fe$_{60}$B$_{20}$(1)/Ti(1) multilayer stacks are characterized by a quite large D parameter of ~1.5 mJ/m$^2$, even if they contain only one heavy metal layer. This is due to the presence of a thin Ti spacer layer which successfully avoids the degradation of i-DMI from symmetric Pt electrodes. Moreover, BLS measurements suggest that the i-DMI strength in the CoFeB/Pt interface is affected by the CoFeB composition. More specifically, we have shown that i-DMI together with dipolar field in these multilayers stabilizes the radial vortex structure for in-plane[24] anisotropy and Néel skyrmions for perpendicular[25] to plane anisotropy.

The magnetic characteristics of the repeated stacks are sensitive to the number of repeats as well as manufacturing conditions. XRD measurements suggest the presence of oxygen in Ti interlayers. Further X-ray Photoelectron Spectroscopy (XPS) studies confirmed partial oxidation of Ti interlayers resulting in predominantly TiO such that the oxide to metal ratio increases with increasing substrate temperature during long hours of deposition (See Fig. S5). Therefore, sample 60C, with higher final substrate temperature of 60 °C compared to that of sample 15C (20°C), has improved interface quality due to decreasing lattice mismatch from 16% to 6% between Ti and Pt which induces a strong Pt <111> orientation and favors strong PMA. Consequently, samples 15C and 60C both have PMA yet due to more oxidation in Ti layers of the latter, PMA is observed to be enhanced.

While PMA is expected to be weaker in sample 15C-AI as well, the post-production treatment of annealing at 170 °C for 45 minutes and ion-irradiation during milling for 2 hours further weakens PMA and switches the easy axis to in plane. A similar effect was observed previously[26] with ion irradiation for the case of Co/Pt. Ion irradiation was reported to cause lattice distortion resulting in an increase in lattice mismatch favoring in-plane anisotropy.

In sample 15PP-A which displays PMA, magnetic structures like Néel skyrmions and horse-shoes appear in the disks with 500 and 1200 nm, respectively. Whereas the in-plane anisotropy in sample 15FP-AI leads to the observation of dot-like features with core radii 50-100 nm in MFM for disks of 300-500 nm diameter. Majority of these dot-like structures responded to in-plane external magnetic field by propagating along the field axis either parallel or anti-parallel to the field direction. Moreover, with a proper external out-of-plane field treatment they were observed to switch their direction of motion. Our results are in-line with the predictions of ref.[3] which indicate the nucleation of radial vortices at the interface of Pt/CoFeB in structures with in-plane anisotropy. Furthermore, we demonstrate that different magnetic treatments can lead to the nucleation of radial vortex structure on the same nanopillar disk with a switched chirality.

The magnetic structures revealed here were nucleated with an external magnetic field treatment but did not require a dc bias-field to be stabilized afterwards which was found to be necessary in some earlier studies[7,27]. Other skyrmion nucleation techniques include the application of external magnetic field pulses[15] or injection of current pulses through a constriction[14,28]. Usually, demonstrated skyrmions are in a single ultra-thin magnetic layer with a few exceptions[7,25].

The Pt/CoFeB/Ti multilayer whose magnetic anisotropy can be tuned with number of repeats and post-production treatments, provides a unique experimental workbench to study the spectrum of micro-magnetic configurations from Néel skyrmions, horse-shoes, spider-web like domains to radial vortices thanks to interfacial DMI and dipolar field. These extraordinary micromagnetic configurations can potentially find applications in magnetic memory, logic and sensor technologies.

The radial vortex state which has been stabilized at room temperature for the first time in this work paves the way to unique applications such as particle trapping/detrapping in addition to nonuniform magnetization memories with efficient switching, where the magnetic layer that accommodates the radial vortex can act as a source of spin-polarized current with radial polarization.

**Methods**

**Sample Preparation**

Samples were fabricated via top-down method utilizing a combination of DC magnetron sputtering (AJA INTERNATIONAL, Inc.) and thermal evaporation (Nanovak, NVTE4-01) for thin film growth. Sample stack was sputtered on $Si/SiO_2$ substrates at room temperature in 2mTorr Ar environment starting from a base pressure on the order of $10^{-9}$ Torr. The sputter deposition was performed using a DC power of 100, 100, 50 Watts resulting in deposition rates of 0.34, 1.07 and 0.18 Å/s, for Ti, Pt and CoFeB respectively. A layer of C was evaporated on top of the film to act as a mask for the pattern. Electron-beam lithography (Vistec, EBPG5000plusES) was used to pattern the disks on a thin layer of e-beam resist spun onto the C layer on top of which Cr film was evaporated. After a lift-off process where the resist was removed in acetone and isopropanol solutions respectively, Cr shaped disks were left on the film-stack. These disk patterns served as a mask to etch off the excess C when exposed to an $O_2$ plasma. The wafers then were placed in an ion-milling system with an ion current density of ~ 0.06 mA/cm$^2$ (corresponding to an approximate fluence on the order of $10^{14}$ ions/cm$^2$) where Cr and especially C which has a very low etch rate (~0.1 Å/s) mask the disk patterns and the rest of the film stack was etched away all the way to the substrate. The remainder C layer was removed by plasma ashing technique.

**Brillouin Light Scattering**

BLS measurements were performed on a single repeat film (1C) and 10 repeat stack (10C) film with monochromatic light from a solid-state laser with wavelength λ=532 nm operating at a power of 100 mW. Sandercock-type (3+3)-pass tandem Fabry-Perot interferometer was used to analyze the back-scattered light. A bias field was applied parallel to the surface plane, while the in-plane wave vector k was swept along the perpendicular direction (Damon-Eshbach configuration). Due to the photon-magnon conservation law of momentum in the scattering process, the magnitude of the wave vector is linked to the incidence angle of light θ, by the relation k=4πsinθ/λ. In our measurements, k was changed from 0 to 2.22 × 10 7 rad/m. The frequency difference, Δ$f$, between Stokes and anti-Stokes peaks is plotted against wave number k for 5 kOe bias field applied parallel to the film plane; k values are inverted for negative bias field -5 kOe. The strength of i-DMI, D has been determined by a linear fit of the experimental data, using equation (1) using gyromagnetic ratio as 190 GHz/T from ref. 17 and $M_s$ measured with bulk CoFeB by VSM as 1185 kA/m.

### Magnetic Characterization

Hysteresis loop measurements of the samples were made using Quantum Design Physical Properties Measurement System (PPMS) Vibrating Sample Magnetometer (VSM) option. Measurements can be performed under either out-of-plane or in-plane magnetic field (up to 9 T). The samples were cut in proper dimensions (∼ 3.5x4 mm$^2$) and mounted on the chip carrier with *Kapton* tape, which was placed into the VSM system. The background magnetic contribution of the silicon substrate and the sample holder was subtracted from the raw data. Magnetic field was scanned between -5000 Oe and 5000 Oe with a rate of 15 Oe/sec at room temperature.

Magnetic Force Microscope (Bruker, Dimension Edge) and its magnetic material (Co/Cr) coated Silicon tips (Bruker, MESP-Low Moment) were used with custom made magnetic stages which were designed to apply an in-plane or out of plane magnetic field to the sample during the experiments. The resonance frequency of the cantilever was ∼70 kHz. The magnetic tip has a diameter of ∼50 nm. For field conditioning treatments, either a GMW, Dipole Electromagnet 5403 or Quantum Design PPMS was used.

### X-Ray Diffraction

X-Ray diffraction (XRD) technique was used to probe the crystal structure of the multilayer stack samples. We used Bruker D8 Advance system which utilizes monochromatic Cu K$_\alpha$ radiation to produce diffraction patterns to acquire the data in a range from 30$^o$ to 50$^o$ on 2θ axis. This system has 0.02$^o$ angle resolution. JCPDS database was used to identify the observed peaks.

**X-Ray Photoelectron Spectroscopy**

X-Ray photoelectron spectroscopy (XPS) technique was used to determine the chemical composition of the samples. For these measurements, we employed Thermo Scientific K-Alpha XPS system which was operated in constant analyzer energy (CAE) mode with 50 eV pass energy, 0.1 eV energy step size for high resolution scans (Fig. S5) and 400 μm focused beam spot size.

CasaXPS software was utilized for calibration, elemental identification and quantification of XPS spectra. Binding energy offsets of XPS spectra due to instrumental errors were corrected using C 1s reference peak located 284.8 eV on the survey scan. This software allowed us to deconvolute the high resolution XPS spectra to unravel chemical properties of samples.

**Acknowledgements**

This work was financially supported by TUBITAK and FEDER, the Region Lorraine and the Grand Nancy and the bilateral agreement France-Turkey (TUBITAK-CNRS) project (TUBITAK Grant # 114F318, CNRS Grant # PICS42452), by the project PRIN2010ECA8P3 from Italian MIUR, by the bilateral agreement Italy-Turkey (TUBITAK-CNR) project (CNR Grant # B52I14002910005, TUBITAK Grant # 113F378). This work was supported partly by the French PIA project "Lorraine Université d'Excellence", reference ANR-15-IDEX-04-LUE.

**Author Contributions**

V.K., A.T.H., S.A., K.O., S.T. designed and performed experiments; V.K., S.A., K.O., I.C. and C.Y. participated in sample preparation; R.T., G.S., M.C., G.F. performed micromagnetic simulations; V.K., A.G., A.T.H., S.A., K.A., S.T., G.F., T.H. and O.O. analyzed data and wrote the manuscript. O.O. supervised the project. All authors contributed to the manuscript.

**Additional Information**

**Competing Interests:** The authors declare that they have no competing interests.